\documentclass{article}
\usepackage{spconf,amsmath,graphicx,amssymb}
\usepackage{mathtools}
\usepackage{siunitx}

\usepackage{balance,cite}

\usepackage{tabularx, etoolbox, booktabs}
\usepackage{multirow} 
\newcolumntype{H}{>{\setbox0=\hbox\bgroup}c<{\egroup}@{}}  
\usepackage{subcaption}
\usepackage{algorithm}

\title{Integrating end-to-end neural and clustering-based diarization: \\ Getting the best of both worlds 
}

\name{Keisuke Kinoshita, Marc Delcroix, Naohiro Tawara}
\address{NTT Corporation, Japan}

\begin{document}
\ninept
\maketitle
\begin{abstract}
Recent diarization technologies can be categorized into two approaches, i.e., clustering and end-to-end neural approaches, which have different pros and cons.
The clustering-based approaches assign speaker labels to speech regions by clustering speaker embeddings such as x-vectors.
While it can be seen as a current state-of-the-art approach that works for various challenging data with reasonable robustness and accuracy,
it has a critical disadvantage that it cannot handle overlapped speech that is inevitable in natural conversational data.
In contrast, the end-to-end neural diarization (EEND), which directly predicts diarization labels using a neural network, 
was devised to handle the overlapped speech.
While the EEND, which can easily incorporate emerging deep-learning technologies, has started outperforming the x-vector clustering approach in some realistic database,
it is difficult to make it work for {\it long} recordings (e.g., recordings longer than 10 minutes) because of, e.g., its huge memory consumption.
Block-wise independent processing is also difficult because it poses an inter-block label permutation problem, i.e., an ambiguity of the speaker label assignments between blocks.
In this paper, we propose a simple but effective hybrid diarization framework that works with overlapped speech 
and for long recordings containing an arbitrary number of speakers.
It modifies the conventional EEND framework to output global speaker embeddings so that speaker clustering can be performed across blocks based on a constrained clustering algorithm to solve the permutation problem.
With experiments based on simulated noisy reverberant 2-speaker meeting-like data, 
we show that the proposed framework works significantly better than the original EEND especially when the input data is long.

\end{abstract}
\section{Introduction}
\label{sec:intro}
Automatic meeting/conversation analysis is one of the essential technologies required 
for realizing futuristic speech applications such as communication agents that can follow, respond to, and facilitate our conversation. 
As an important central task for the meeting analysis, speaker diarization has been extensively studied \cite{Diarization_review, DIHARD_data, AMI_data}.

Current state-of-the-art diarization systems that achieve reliable performance in many challenges \cite{Diarization_review, DIHARD_data} is based on clustering of speaker embeddings (i.e., speaker identity features) such as i-vectors \cite{i-vector} and x-vectors \cite{x-vector}.
Such clustering-based approaches first segment a recording into short homogeneous blocks and compute speaker embeddings for each block 
assuming that only one speaker is active in each block. 
Then, speaker embedding vectors are clustered to regroup segments belonging to the same speakers and obtain the diarization results. 
Various speaker embeddings and clustering techniques have been explored in \cite{DIHARD_JHU,DIHARD_BUT,Zhang_ICASSP19, Li_2020_RelationNet}.
While these methods can cope with very challenging scenarios \cite{DIHARD_JHU,DIHARD_BUT}
and work with an arbitrary number of speakers,
there is a clear disadvantage that they cannot handle overlapped speech,
i.e., time segments where more than one person is speaking, because of the way of extracting speaker embeddings. 
Perhaps surprisingly, even in professional meetings, the percentage of overlapped speech is in the order of 5 to 10\%, 
while in informal get-togethers it can easily exceed 20\% \cite{onlineRSAN_ICASSP2019}.

End-to-End Neural Diarization (EEND) has been recently developed \cite{Fujita_IS2019,Fujita_ASRU2019,Horiguchi2020_EDA_EEND} to address the overlapped speech problem.
Similarly to the neural source separation algorithms \cite{Kolbaek2017,RSAN},
in EEND, a Neural Network (NN) receives standard frame-level spectral features and directly outputs a frame-level speaker activity
for each speaker, 
no matter whether the input signal contains overlapped speech or not.
While the system is simple and has started outperforming the conventional clustering-based algorithms \cite{Fujita_ASRU2019,Horiguchi2020_EDA_EEND},
it is difficult to directly apply the EEND systems to {\it long} recordings (e.g., recordings longer than 10 minutes).
The system is designed to operate in a batch processing mode and thus requires a very large  computer memory when performing inference with long recordings.
Besides, aside from the memory issue, the NNs in EEND has difficulty to generalize to unseen very long sequential data,
which also hampers its application to the long recordings.
Note that, if we segment the long recordings into small chunks and apply the original EEND model to each chunk independently,
the model inevitably suffers from the inter-block label permutation problem,
i.e., an ambiguity of the speaker label assignments between chunks.
To address this problem (and simultaneously seek for a low-latency solution), 
\cite{Xue2020_speaker_tracing} proposed an NN-based extension of the EEND to block-online processing.
The method in \cite{Xue2020_speaker_tracing} first tries to find single speaker regions,
and use them as a guide to assign the speaker labels to the diarization results of future blocks. 
However, their performance typically does not reach that of the original EEND. 
Also, more importantly, the method cannot handle an arbitrary number of speakers.

In this paper, we propose a simple but effective hybrid diarization approach,
called EEND-vector clustering, by combining the best of the clustering-based diarization and the EEND.
A central component of the proposed approach is a modified EEND network that outputs, in each chunk, 
not only the diarization results but also global speaker embeddings 
associated with the diarization results. The inter-block permutation ambiguity problem can thus be simply solved 
by clustering the block-level speaker embedding vectors. 
This extension thus naturally allows us to combine the advantages of both clustering and the EEND based methods, 
i.e. it can work with overlapped speech and deal with long recordings including an arbitrary number of speakers. 
In particular, we confirm experimentally that the proposed EEND-vector clustering significantly outperforms 
the original EEND system especially when the recordings are long, e.g., more than 5 minutes,
while maintaining the same performance as the original EEND system when the recording is short. 

The remainder of this paper is organized as follows. We first introduce the proposed framework in section 2 in detail.
Then, in section 3, we evaluate its performance in comparison with the original EEND to clarify the advantages of the proposed framework.
Finally, we conclude the paper in section 4.


\begin{figure}[t]
 \begin{center}
  \includegraphics[width=85mm]{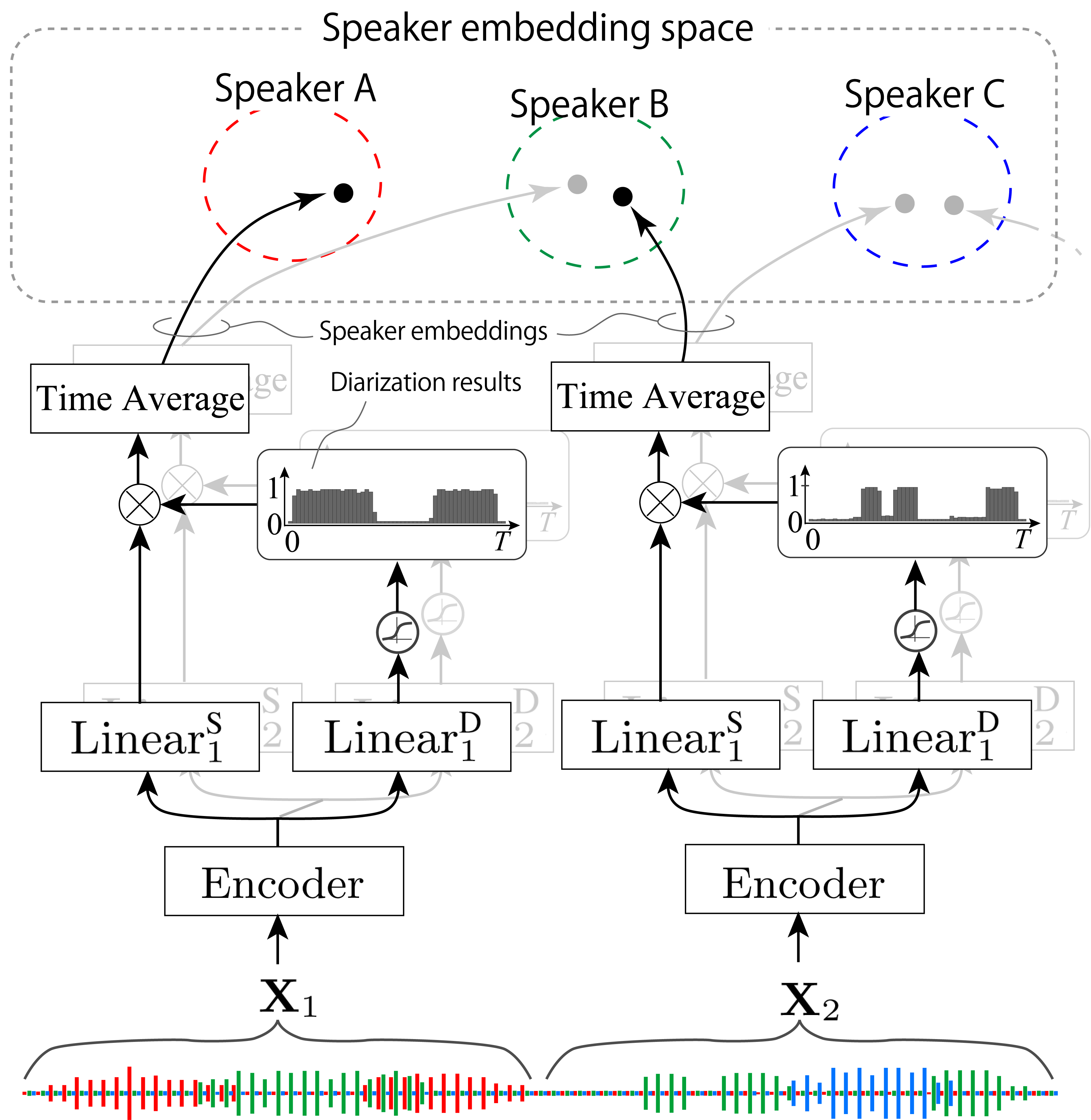}
     \end{center}
   \vspace{-3mm}
   \caption{Schematic diagram of the proposed diarization framework. The input contains 3 speakers in total (red, green, and blue speakers shown in the waveform in the bottom), but only at most 2 speakers are actively speaking in each chunk.}
 \label{fig:overview}
\end{figure}

\section{Proposed Diarization framework: \\EEND-vector clustering}
\label{sec:proposed_method}

\subsection{Overall framework}
\label{sec:framework}
Figure~\ref{fig:overview} shows a schematic diagram of the proposed EEND-vector clustering framework.

It first segments the input recording into chunks 
and calculates a sequence of the input frame features within each chunk, 
as $\mathbf{X}_i = (\mathbf{x}_{t,i} \mid t=1,\cdots,T)$ where $i$,$t$ and $T$ are the chunk index, the frame index in the chunk and the chunk size\footnote{The chunk size $T$ for estimating speaker embeddings can be advantageously much longer than the homogeneous blocks used in x-vector clustering since we can handle {\it heterogeneous} chunks including more than 1 speaker.}.
$\mathbf{x}_{t,i} \in \mathbb{R}^{K}$ is the $K$-dimensional input frame feature at the time frame $t$.
In the example shown in Fig~\ref{fig:overview}, the input recording consists of 2 chunks and contains 3 speakers in total. 
In the following, we assume that we can fix the maximum number of active speakers in a chunk, $S_{\textrm{Local}}$, to 2, 
although the method could be generalized to more speakers or an unknown number of speakers~\cite{Horiguchi2020_EDA_EEND}
\footnote{If we select the chunk size carefully, it is not too difficult to set an appropriate maximum number of speakers even for practical use cases \cite{Yoshioka_ICASSP2019}.}.

Based on the hyper-parameter $S_{\textrm{Local}}=2$,
the network estimates diarization results for 2 speakers in each chunk.
In Fig.~\ref{fig:overview}, the processing for the 1st speaker is drawn with black lines and put in the foreground, 
while that of the 2nd speaker is drawn with grey lines and put in the background.
The diarization results are estimated independently in each chunk through NNs denoted as $\mathrm{Encoder}$ and $\mathrm{Linear}_s^{\textrm{D}}\ (s=1,2)$, where $s$ is the speaker index within a chunk.
Since it is {\it not} always guaranteed that the diarization results of a certain speaker are estimated at the same output node,
we may have the inter-block label permutation problem in the diarization outputs.
As an example, in Fig.~\ref{fig:overview}, the network $\mathrm{Linear}_1^{\textrm{D}}$ estimates the diarization result of `speaker A' in the first chunk,
and that of `speaker B' in the second chunk.
This means that we cannot obtain an optimal diarization result simply by stitching the diarization results of a specific output node across all the chunks.

To solve this permutation problem, we simultaneously estimate a speaker embedding corresponding
to each diarization result in each chunk.
The network to estimate the speaker embeddings are denoted as $\mathrm{Linear}_s^{\textrm{S}} \ (s=1,2)$ in Fig.~\ref{fig:overview}.
The speaker embedding extraction network is optimized through the NN training such that the vectors of the same speaker stay close to each other,
while the vectors of different speakers lie far away from each other.
This can be seen in the figure by examining how the embeddings are organized in the speaker embedding space. 
Therefore, after obtaining diarization results for all chunks,
by clustering the speaker embeddings given the total number of speakers in the input recording (3 in this case),
we can estimate the correct association of the diarization results among chunks.
Then, finally, the overall diarization results are obtained by stitching them together based on the embedding clustering result.
Note that while the proposed framework estimates the diarization results of the fixed number of speakers in a chunk, 
it can handle a meeting with an arbitrary number of speakers.

For the clustering, we can use any clustering algorithms.
However, it may be preferable if the clustering algorithm is aware of the characteristic of this framework 
and work with a constraint that the speaker embeddings from a chunk should not belong to the same speaker cluster.
In this paper, to incorporate the constraint into the clustering stage,
we use a {\it constrained} clustering algorithm called COP-k-means \cite{COP-kmeans},
which allows us to set cannot-link constraints between a given pair of embeddings to prevent the pair from being assigned to the same speaker cluster.

\subsection{Neural diarization with speaker embedding estimation}
\vspace{-1mm}
This subsection details the NN model in EEND-vector clustering to estimate the diarization results and the speaker embeddings.

Let us denote the ground-truth diarization label sequence as $\mathbf{Y}_i = (\mathbf{y}_{t,i} \mid t=1,\cdots,T)$ that corresponds to $\mathbf{X}_i$.
Here, the diarization label $\mathbf{y}_{t,i} = [y_{t,i,s} \in \{0,1\} \mid s=1, \cdots, S_{\textrm{Local}}]$ represents a joint activity for $S_{\textrm{Local}}$ speakers.
For example, $y_{t,i,s} = y_{t,i,s'} = 1 (s \ne s')$ indicates both speakers $s$ and $s'$ spoke at the time frame $t$ in the chunk $i$. 

In the EEND framework, the diarization task is formulated as a multi-label classification problem.
Specifically, we estimate the dirarization result of the $s$-th speaker at each time frame, $\hat{y}_{t,i,s}$, as,
\begin{eqnarray}
    \bigl[\mathbf{h}_{1,i},\ldots,\mathbf{h}_{T,i} \bigr] &=& \mathrm{Encoder}( \mathbf{X}_i ) \in \mathbb{R}^{D \times T}, \nonumber \\
    \hat{y}_{t,i,s} &=& \mathrm{sigmoid}(\mathrm{Linear}_s^{\textrm{D}}(\mathbf{h}_{t,i})) \in (0,1) \nonumber \\ 
                      && \ \ \ \ \ \ \ \ \ \ \ \ \ \ \ \ \ \ \ \ \ \ \ \ \ \ \ (s=1,\ldots,S_{\textrm{Local}}), \label{eq:diarization_out}
\end{eqnarray}
where $\mathrm{Encoder}(\cdot)$ is an encoder such as a multi-head self-attention NN \cite{Fujita_ASRU2019},
which utilizes all the input features $\mathbf{X}_i$ for inference.
$\mathbf{h}_{t,i}$ is a $D$-dimensional internal representation in the NN, 
$\mathrm{Linear}_s^{\textrm{D}}(\cdot) : \mathbb{R}^{D} \rightarrow \mathbb{R}^{1}$ is a fully-connected layer to estimate the diarization result,
and $\mathrm{sigmoid}(\cdot)$ is the element-wise sigmoid function.

Now, after estimating the diarization results, for the purpose of solving the inter-block permutation problem,
we estimate the speaker embedding, $\hat{\mathbf{e}}_{i,s}$, corresponding to the diarization result of the $s$-th speaker as follows.
\begin{eqnarray}
    \mathbf{z}_{t,i,s} &=& \mathrm{Linear}_s^{\textrm{S}}(\mathbf{h}_{t,i}) \in \mathbb{R}^{C}, \nonumber \\
    \bar{\mathbf{z}}_{i,s} &=& \sum_{t=1}^{T} \hat{y}_{t,i,s} \mathbf{z}_{t,i,s}, \in \mathbb{R}^{C} \label{eq:weighted_sum} \\
    \hat{\mathbf{e}}_{i,s} &=& \frac{\bar{\mathbf{z}}_{i,s}}{\| \bar{\mathbf{z}}_{i,s} \|} \in \mathbb{R}^C \ \ \ (s=1,\dots,S_{\textrm{Local}}), \label{eq:speaker_embedding}
\end{eqnarray}
where $C$ is the dimension of the speaker embedding, $\mathrm{Linear}_s^{\textrm{S}}(\cdot) : \mathbb{R}^{D} \rightarrow \mathbb{R}^{C}$ is a fully-connected layer to estimate the $s$-th speaker's embedding $\mathbf{e}_{i,s}$,
and $\|\cdot\|$ is a vector norm.
Here we chose to estimate the speaker embeddings as weighted sum of frame-level embeddings $\mathbf{z}_{t,i,s}$ 
with weights determined by the diarization results $\hat{y}_{t,i,s}$, as in Eq.~(\ref{eq:weighted_sum}).
With these operations, we can estimate diarization results and speaker embeddings for all $S_{\textrm{Local}}$ speakers.
This model without the speaker embedding estimator is essentially the same as the conventional EEND~\cite{Fujita_ASRU2019}.

\subsection{Training objectives}
Now, we will explain a way to train the model to realize the behavior explained in Section~\ref{sec:framework}.
Since the network estimates both the diarization results and speaker embeddings simultaneously,
our natural choice is to use the following multi-task loss.
\begin{eqnarray}
\mathcal{L} &=& (1-\lambda) \mathcal{L}_{\textrm{diarization}} + \lambda \mathcal{L}_{\textrm{speaker}}, \label{eq:total_loss} 
\end{eqnarray}
where $\mathcal{L}$ is the total loss function to be minimized, $\mathcal{L}_{\textrm{diarization}}$ is the diarization error loss,
$\mathcal{L}_{\textrm{speaker}}$ is speaker embedding loss, and $\lambda$ is a hyper-parameter to weight the two loss functions.

\subsubsection{Diarization loss}
Following \cite{Fujita_IS2019}, the diariation loss in each chunk is formulated as:
\begin{eqnarray}
\mathcal{L}_{\textrm{diarization},i}, \phi^{\star} &=& \frac{1}{TS_{\textrm{Local}}} \min_{\phi \in \mathrm{perm}(S_{\textrm{Local}})} \sum_{t=1}^{T} \textrm{BCE}\left( \mathbf{l}_{t,i}^{\phi},\hat{\mathbf{y}}_{t,i}  \right), \label{eq:diarization_loss} 
\end{eqnarray}
where $\mathrm{perm}(S_{\textrm{Local}})$ is the set of all the possible permutations of ($1,\dots,S_{\textrm{Local}}$),
$\hat{\mathbf{y}}_{t,i}=[\hat{y}_{t,i,1},\ldots,\hat{y}_{t,i,S_{\textrm{Local}}}] \in \mathbb{R}^{S_{\textrm{Local}}}$,
$\mathbf{l}_{t,i}^{\phi}$ is the $\phi$-th permutation of the reference speaker labels, and $\mathrm{BCE}(\cdot, \cdot)$ is the binary cross-entropy function between the labels and the estimated diarization outputs.
$\phi^{\star}$ is the permutation that minimizes the right hand side of the Eq.~(\ref{eq:diarization_loss}).
This training scheme called permutation-invariant training has shown to be effective for the neural diarization \cite{Fujita_IS2019}, 
but at the same time, it incurs another problem, i.e., the inter-block label permutation problem since it clearly allows the speaker labels to permute from chunk to chunk.
The diarization loss function $\mathcal{L}_{\textrm{diarization}}$ is formed by collecting $B$ chunks, i.e., $\mathcal{L}_{\textrm{diarization}}=\sum_{i=1}^{B} \mathcal{L}_{\textrm{diarization},i}$, where $B$ is the size of the mini-batch.

Here, as it was mentioned earlier, 
$S_{\textrm{Local}}$ is a hyper-parameter that has to be appropriately chosen to satisfy (1) $S_{\textrm{Local}} \leq S_{\textrm{total}}$
where $S_{\textrm{total}}$ is the total number of speakers in the recording, and (2) $S_{\textrm{Local}}$ is always greater than or equal to the maximum number of speakers speaking in a chunk.
With an assumption that $S_{\textrm{Local}}$ is chosen in such a way, 
the diarization labels in the chunk $i$, $\mathbf{Y}_i$, should be formed 
as a subset of all $S_{\textrm{total}}$ speaker's labels $\mathbf{Y}_i^{\mathrm{total}}$, 
i.e., $\mathbf{Y}_i \subseteq \mathbf{Y}_i^{\mathrm{total}}$.
The subset should be chosen appropriately for each chunk such that it covers all speakers speaking in the chunk $i$.
If the number of speakers speaking in the chunk is smaller than $S_{\textrm{Local}}$,
we fill $\mathbf{Y}_{i}$ with diarization label(s) of a virtual $(S_{\textrm{total}}+1)$-th always-silent speaker, 
i.e., $(y_{t,i,S_{\textrm{total}}+1} \in \{0\} \mid t=1,\ldots,T)$.

\subsubsection{Speaker embedding loss}
For the speaker embedding training, 
we use a loss function that encourages the embeddings to have small intra-speaker and large inter-speaker distances.
Specifically, we utilize the loss proposed recently in \cite{Wavesplit} , which was shown to be very effective for the speech separation task.
For this loss function, we assume that the training data is annotated with speaker identity labels, i.e., indices, based on a finite set of $M$ training speakers.
Note, however, that the speaker identity is not required at test time, and that training and test speakers can differ (i.e., open speaker conditions).
Let $\sigma_i^{\star} = \bigl[ \sigma_{i,1}^{\star},\ldots,\sigma^{\star}_{i,S_{\textrm{Local}}} \bigr]$ 
be the absolute speaker identity indices that correspond to the permutation of the labels that gives minimum value to Eq.~(\ref{eq:diarization_loss}), i.e., $\phi^{\star}$.
$\sigma^{\star}_i$ is a subset of the $M$ speaker identity indices.
Then, the speaker embedding loss for chunk $i$, $\mathcal{L}_{\textrm{speaker},i}$, is formulated as follows.
\begin{eqnarray}
\mathcal{L}_{\textrm{speaker},i} &=&  \frac{1}{S_{\textrm{Local}}} \sum_{s=1}^{S_{\textrm{Local}}} l_{\textrm{speaker}} \left(\sigma^{\star}_{i,s}, \hat{\mathbf{e}}_{i,s} \right), \label{eq:speaker_loss_1} 
\end{eqnarray}
where
\begin{eqnarray}
l_{\textrm{speaker}} \left(\sigma^{\star}_{i,s}, \hat{\mathbf{e}}_{i,s} \right) &=&   
    - \ln \left( \frac{\exp{\left( -d\left(E_{\sigma^{\star}_{i,s}}, \hat{\mathbf{e}}_{i,s} \right) \right)}}{ \sum_{m=1}^M \exp{\left( -d\left(E_{m}, \hat{\mathbf{e}}_{i,s} \right) \right)}} \right), \label{eq:speaker_loss_2} \\
d \left(E_{m}, \hat{\mathbf{e}}_{i,s} \right) &=& \alpha \| E_{m} - \hat{\mathbf{e}}_{i,s} \|^2 + \beta, \label{eq:speaker_loss_3}
\end{eqnarray}
where $E$ is a learnable global speaker embedding dictionary, 
and $E_{m}$ is a learnable variance-normalized global
speaker embedding associated with the $m$-th training speaker.
Eq.~(\ref{eq:speaker_loss_3}) is the squared Euclidean distance between the learnable global speaker embedding
and the estimated speaker embedding, which is rescaled with learnable scalar parameters $\alpha>0$ and $\beta$.
Eq.~(\ref{eq:speaker_loss_2}) is the log softmax over the distances between the estimated embedding and the global embeddings,
which can be derived from the categorical cross-entropy loss.
The loss function $\mathcal{L}_{\textrm{speaker}}$ is formed by collecting $B$ chunks, similarly to $\mathcal{L}_{\textrm{diarization}}$.

By minimizing these loss functions, we expect to estimate diarization results accurately even if there is overlapped speech,
and simultaneously estimate speaker embeddings that are suitable for the subsequent clustering process.



\section{Experiments}
In this section, we evaluate the effectiveness of the proposed method in comparison with the conventional EEND \cite{Fujita_ASRU2019},
based on test data including long recordings with a significant amount of overlapped speech.
Comparison with the x-vector clustering is omitted since it was already shown in \cite{Fujita_ASRU2019} 
that the conventional EEND works better in case the data contains overlapped speech.

\subsection{Data}
The training, development, and test data are based on the 8~kHz Librispeech database \cite{Librispeech}.
To simulate a conversation-like mixture of two speakers, 
we picked up utterances from randomly selected two speakers,
and generated a noisy reverberant mixture containing many utterances per speaker 
with reasonable silence intervals between utterances. 
For the simulation, we used the algorithm proposed in \cite{Fujita_IS2019},
and set the average silence interval between utterances at 2 seconds.
Noise data was obtained from MUSAN noise data~\cite{MUSAN}.
The signal-to-noise ratio was sampled randomly for each mixture from 5, 10, 15, and 20~dBs.
For reverberation, we used 20000 impulse response data in~\cite{Ko_2017}, which simulates various rooms.
Consequently, we obtained a set of training, development, and test data that contains various overlapping ratios 
ranging from 10 to 90~\%.

For the training and development data, 
we randomly selected utterances from 460-hour clean speech training data containing 1172 speakers ($M=$1172)
and generated 40000 and 500 mixtures that amount to 2774 and 23 hours, respectively.
For the test data, we generated 4 different sets of data that differ in duration.
Each test set contains 500 utterances. 
The average duration of mixtures in each set is 3, 5, 10, and 20 minutes, respectively.
All the test data were generated based on the Librispeech test set containing 26 speakers 
that were not included in the training and development data.

\begin{table}[t]
\centering
\caption{DERs (\%) of the conventional EEND and the proposed models for each test set that differs in the duration.}
\vspace{-3mm}
\label{tbl:overall_result}
\scalebox{0.95}[0.95]{
\begin{tabular}{l c c c c c c}
\toprule
Model     & Chunking & Clustering &  \multicolumn{4}{c}{Test data duration (minutes)}  \\ \cline{4-7}
&            &          &   3        &  5                &  10  & 20 \\ 
\midrule
1. EEND  &     -  &    N/A   &   8.0 &  8.7  &  9.3  & N/A     \\ 
2. EEND  & \checkmark&  N/A &   10.1        &  10.0                 &  10.3                 & 10.1    \\
\midrule
3. Proposed  &  -     &  -   &   \textbf{7.4}     &  8.5 &  9.2 & N/A      \\ 
4. Proposed  & \checkmark &  -    &   9.3    &  9.3   &  9.6 & 9.4     \\ 
5. Proposed  & \checkmark &  \checkmark &   \textbf{7.4}  &  \textbf{6.5}    &  \textbf{5.9} & \textbf{5.5}\\ 
\bottomrule
\end{tabular}
}
\end{table}

\begin{table}[t]
\centering
\caption{DERs (\%) of the conventional EEND and the proposed EEND-vector clustering for each overlap condition.}
\vspace{-3mm}
\label{tbl:result_overlap_ratio}
\scalebox{0.97}[1.0]{
\begin{tabular}{l c c c c c }
\toprule
Model     & Chunking  &  Clustering  &   \multicolumn{3}{c}{Overlap ratio (\%) }  \\ \cline{4-6}
          &            &              &  0 - 30     &  30 - 60          &  60 - 90  \\ 
\midrule
EEND      &     -       &     -       &   9.8        &  9.6                &  7.2          \\ 
Proposed  & \checkmark  &  \checkmark & \textbf{4.6}  &  \textbf{6.0}       &  \textbf{5.7} \\ 
\bottomrule
\end{tabular}
}
\end{table}

\subsection{NN training and hyper-parameters}
\vspace{-0mm}
For the input frame feature, we extracted 23-dimensional log-Mel-filterbank features 
with 25~ms frame length and 10~ms frame shift. 

For both the proposed method and the conventional EEND,
the chunk size $T$ at the training stage was set at 500 ($=$ 50 seconds) as in \cite{Fujita_ASRU2019}.
Therefore, when the training data is longer than 50 seconds,
we split the input audio into non-overlapping 50-second chunks. 
At the inference stage, the conventional EEND uses an entire sequence for inference without chunking.
On the other hand, the proposed method segments the input data into 50-second non-overlapping chunks,
and perform diarization as explained in Section \ref{sec:framework}.
About 25~\% of the chunks in the test data contain less than 2 speakers, while the other contain 2 speakers.

For both methods, we used the same network architecture as \cite{Fujita_ASRU2019}.
For $\mathrm{Encoder}$, we used two multi-head attention blocks 
with 256 attention units containing four heads ($D = 256$).
We used the Adam optimizer with the learning rate scheduler introduced in \cite{Transformer_Vaswani}.
The number of warm-up steps used in the learning rate scheduler was 25000. 
The batch size $B$ was 64. The number of training epochs was 70. 
The final models were obtained by averaging the model parameters of the last 10 epochs. 

For the proposed method, $\lambda$ was set at $0.01$.
With an assumption that the maximum number of speakers speaking in each chunk is 2 or less,
we set $S_{\textrm{Local}}$ at 2.
The dimension of the speaker embedding, $C$, was set at 256.
Since the performance of the proposed method slightly changes due to the initialization of the COP-k-means algorithm, 
we ran the test inference 10 times with random initialization and obtained the averaged results. 
The standard deviation of the obtained diarization error rate (DER) was less than 0.2\%.

\begin{figure}[t]
 \begin{center}
  \includegraphics[width=70mm]{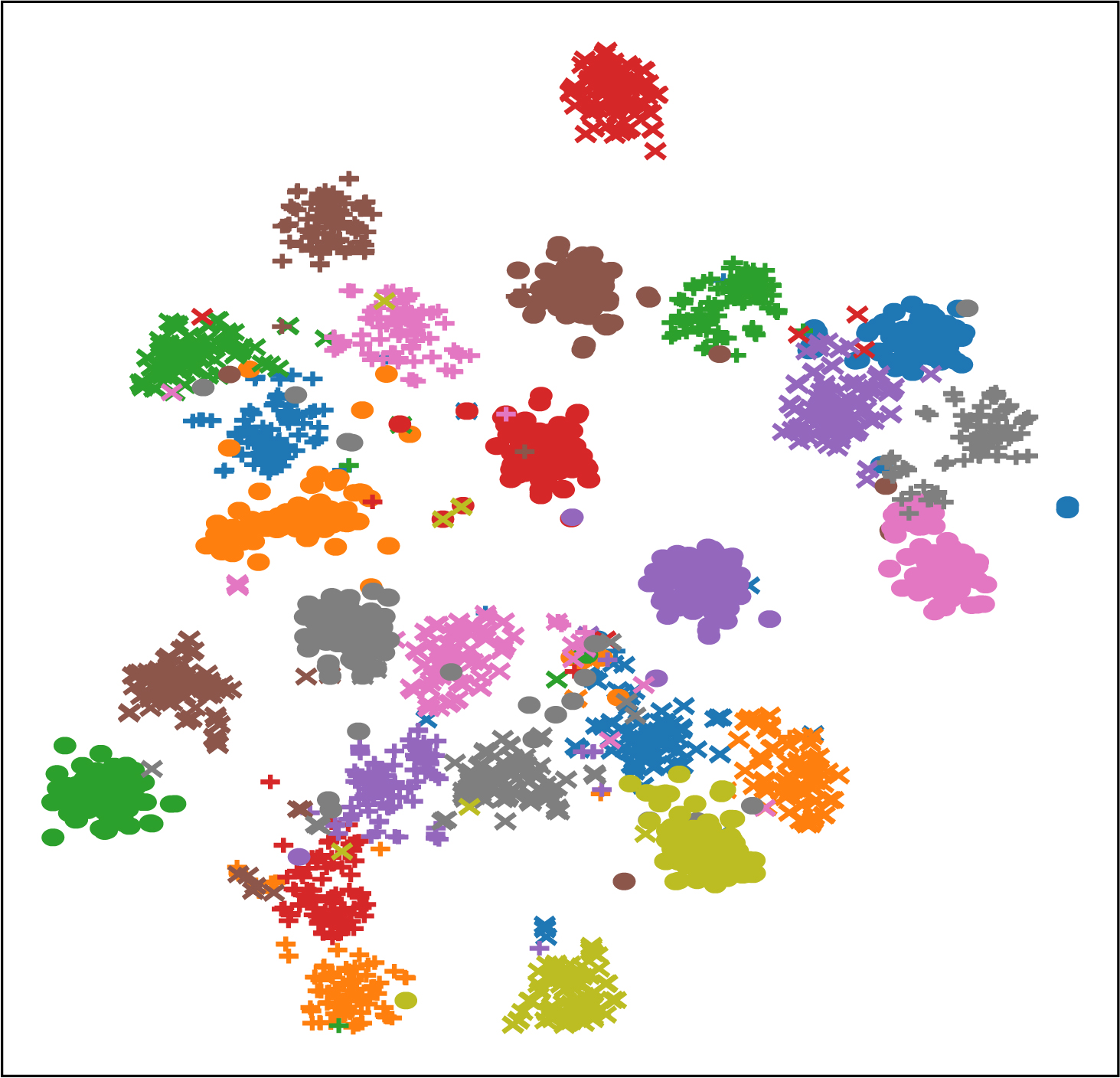}
     \end{center}
  \vspace{-3mm}
  \caption{t-SNE plot of the 26 test speaker's embedding vectors}
 \label{fig:speaker_embedding}
\end{figure}

\subsection{Results}
\vspace{-2mm}
Table~\ref{tbl:overall_result} shows the results of the conventional EEND (1st row) and the proposed method (5th row).
The table contains some variants of these methods to clarify the effectiveness of each component in the proposed model.

First, by comparing the 1st row (conventional EEND applied to the entire sequence without chunking) 
and 5th row (the proposed model that processes chunks and performs clustering, i.e., EEND-vector clustering), 
we can see that the proposed method outperforms the conventional EEND.
In particular, as the duration of the test data gets longer, the proposed method becomes increasingly advantageous.
While the conventional EEND cannot well handle 10- and 20-minute data because of poor generalization to the long data 
and the CPU memory constraint,
EEND-vector clustering can achieve stable diarization performance for such data.
Interestingly, it tends to work better (at least for this data) especially when the duration of the data is long.
It is probably because the number of embeddings available for the clustering becomes larger as the data gets longer,
which helps the clustering algorithm find better cluster centroids.

Now, let us compare the 1st row (EEND without chunking) and 3rd row (the proposed model applied to the entire sequence without chunking).
The performance of the proposed model turned out to be almost equal to that of the conventional method in all cases, 
which indicates that the additional speaker loss did not negatively affect the diarization capability of the model.
The results show that the additional speaker loss did not negatively affect the diarization capability of the model.

Next, let us focus on the comparison between 1st/3rd rows (models without chunking) and 2nd/4th rows (models with chunking but without clustering). 
The performance degradation when using chunking reveals the inter-block label permutation problem.
We assume this problem may become even more severe when dealing with more speakers.
With this comparison, we could confirm the effectiveness of the proposed constrained-clustering-based diarization result stitching.

\subsection{Detailed analysis}
\vspace{-1mm}
\subsubsection{Evaluation in terms of overlapping ratio}
Table~\ref{tbl:result_overlap_ratio} shows the DERs in each overlap condition.
The results were obtained from the test set of 10-minute mixtures.
Since each mixture in the test set differs in the amount of overlapped speech, i.e., overlap ratio,
we categorized the mixtures into several overlap ratio ranges and obtained DER in each condition.
, to better understand the model behavior.
The proposed method is shown to largely outperform the conventional EEND in all conditions.

\subsubsection{Speaker embedding estimation accuracy}
Here we also examine whether the speaker embeddings of the test data is estimated accurately 
such that they have large inter-speaker and small intra-speaker distances.
Figure~\ref{fig:speaker_embedding} shows the t-SNE visualization of the speaker embeddings of the 26 test speakers,
which are obtained from the 10-minute test data. 
It clearly shows distinguished clusters for each speaker,
which proves that we can estimate the {\it global} speaker embeddings accurately
even if the input data contains a significant amount of overlapped speech.

\section{Conclusions}
\vspace{-2mm}
We proposed a simple but effective diarization framework, EEND-vector clustering, that estimates both diarization results and speaker embeddings.
By utilizing the speaker embeddings, we solved the inter-block label permutation problem.
Experimental results showed that EEND-vector clustering works significantly better than the original EEND especially when the input data is long.
Future work includes application of the proposed framework to more challenging conditions as well as an extension to a scheme that can handle an arbitrary number of speakers within a chunk, e.g., \cite{Horiguchi2020_EDA_EEND}.

\bibliographystyle{IEEEbib}
\bibliography{main}

\end{document}